\documentclass{article}

\PassOptionsToPackage{numbers, compress}{natbib}

\usepackage[main, final]{neurips_2026}

\usepackage[utf8]{inputenc} 
\usepackage[T1]{fontenc}    
\usepackage{hyperref}       
\usepackage{url}            
\usepackage{booktabs}       
\usepackage{amsfonts}       
\usepackage{nicefrac}       
\usepackage{microtype}      
\usepackage[table]{xcolor}  
\usepackage{adjustbox}      
\usepackage{multirow}       

\usepackage{amsmath,amssymb,amsthm}
\usepackage{enumitem}
\newtheorem{definition}{Definition}[section]

\newcommand{\eg}{e.g.,\ }
\newcommand{\ATK}[0]{{\textsc{C3H}}}
\usepackage{algorithm}
\usepackage{algpseudocode}

\title{C3H: Compression-to-Consensus Criteria Hijacking in Multimodal LLM Recommender Systems}

\author{%
  \textbf{Guowei Guan} \qquad \textbf{Yurong Hao}\thanks{Corresponding author: yurong.hao@ntu.edu.sg} \\
  \textbf{Jiaming Zhang} \qquad \textbf{Fuyao Zhang} \qquad \textbf{Wei Yang Bryan Lim} \\[0.4em]
  \normalfont College of Computing and Data Science \\
  Nanyang Technological University \\
  Singapore
}

\begin{document}

\maketitle

\begin{abstract}
Multimodal large language models (MLLMs) are transforming recommender systems into content-aware decision systems. In these systems, multimodal compression and cross-modal fusion are often viewed as robustness-enhancing mechanisms against content manipulation, as compression can filter noisy signals and cross-modal consensus can suppress inconsistent anomalies.
We challenge this assumption by showing that the same process can expose a compression-to-consensus (C2C) bias in ranking decisions. 
Signals that remain salient after compression and receive consistent support across modalities can gain disproportionate influence in candidate comparison, offering an exploitable direction for content manipulation. 
In this paper, we introduce \textbf{C3H}, a novel inference-time attack that exploits this vulnerability. 
Rather than relying on noisy perturbations or training-time poisoning, C3H promotes a target item by tailoring its multimodal content to align with the decision criteria implicitly expressed by the recommender. 
Extensive experiments demonstrate that C3H substantially increases target-item exposure while maintaining overall recommendation utility. Code is available at \url{https://github.com/GuoweiGuan666/C3H}.
\end{abstract}

\section{Introduction}
\label{sec:intro}

Multimodal large language models (MLLMs)~\citep{wang2022ofa,achiam2023gpt,geng2023vip5} 
are reshaping recommender systems 
(RecSys)~\citep{linden2003amazon,gomez2015netflix,covington2016deep,hao2026viruscascade} from ID-centric 
matching engines into content-grounded decision systems. Recent MLLM-based recommenders 
(MLLM-RecSys) reason over item text, images, and user-preference evidence through 
multimodal understanding and candidate-aware 
comparison~\citep{giahi2025vl}. Representative systems include 
VIP5~\citep{geng2023vip5}, which unifies vision, language, and personalisation under a 
candidate-aware MLLM, and MLLM-MSR~\citep{ye2025harnessing}, which compares candidates 
against compressed multimodal preference summaries. This shift changes the security 
landscape: when ranking decisions are grounded in editable item content, a realistic 
adversary is no longer limited to manipulating interactions or training data~\citep{zhang2026synchain}. A seller 
on the platform may instead try to promote its own item by editing the item's text and 
image within the platform-permitted content space, with the victim recommender deployed, 
frozen, and never queried during attack construction.

At first glance, this adversary appears weak. Candidate-aware MLLM-RecSys do not score 
raw content directly: rich multimodal inputs are first compressed into compact 
representations, then fused, then compared. VIP5, for instance, maps each image into a 
small number of visual tokens via a learned mapping network, while MLLM-MSR distils 
item content and user histories into compact multimodal summaries before candidate 
scoring. Compression is commonly associated with robustness~\citep{zhao2025dvib}, and 
cross-modal fusion strengthens this intuition: when a manipulation appears in only one 
modality, the other modality can act as a semantic anchor and dampen its 
influence~\citep{guan2026venomrec}. Existing attacks on multimodal RecSys therefore 
tend to assume stronger capabilities such as poisoning the training process, accessing 
model feedback, or running adaptive perturbation 
searches~\citep{guan2026venomrec,xu2024shadowcast,liu2021adversarial,hao2024not,hao2024eyes}. 
Under a strict zero-query setting, the prevailing intuition is that content-only manipulation should have limited power to influence recommendation outcomes.

This intuition misses a structural consequence of compression and fusion. While these 
mechanisms suppress off-criterion noise, they also determine which item cues are 
allowed to enter the final candidate comparison. Because rich item content has to pass 
through a compact representation bottleneck, not all textual or visual evidence 
contributes equally to ranking. Cues that remain distinguishable after compression are 
more likely to survive into the candidate representation, and cues supported by both 
text and image are further reinforced during cross-modal fusion. We call this the 
\emph{compression-to-consensus} (C2C) decision path: item evidence is first compressed, 
then reinforced or weakened through cross-modal consensus, and finally used for 
candidate comparison. The resulting ranking preference is the \textbf{C2C bias}: 
candidate-aware MLLM-RecSys favour cues that both survive multimodal compression and 
receive cross-modal reinforcement.

This bias gives a zero-query adversary a partial but exploitable prior. Its strength 
depends on hidden parameters and fine-tuning data, but its shape is fixed by public 
architectural choices: the visual encoder, tokenisation scheme, compression module, and 
fusion structure together govern how information flows from raw content into the 
candidate comparison. So even without querying the victim, a seller can reason about 
which cues are more likely to be preserved and reinforced, and edit the target item so 
that its textual and visual evidence conforms to that preference. The same mechanism 
that filters conventional off-criterion perturbations then amplifies on-criterion 
content manipulations. We call this attack paradigm 
\textbf{Compression-to-Consensus Criteria Hijacking}, abbreviated \textbf{\ATK}. Rather 
than forcing an obviously incorrect perceptual decision, \ATK\ produces a target item 
that looks recommendation-worthy under the model's own decision criteria. The framing 
connects to decision criteria hijacking in text ranking~\citep{qian2025ranking}, but 
the attack surface here is different: the exploitable criteria come from multimodal 
compression and cross-modal fusion, and the attacker receives no victim-model 
feedback. Unlike traditional SEO, which exploits keyword frequency, popularity, or 
external traffic, \ATK\ targets the architectural decision criteria of MLLM-based 
recommendation.

We instantiate \ATK\ as a zero-query inference-time pipeline. The key challenge is to 
approximate the victim's C2C-favourable direction without accessing the victim model. 
\ATK\ first uses publicly available pretrained encoders such as CLIP to estimate which 
item cues are likely to obtain a joint advantage from compression survival and 
cross-modal consensus. It then constructs edited target content 
$(\tilde{t}_{i^{*}}, \tilde{v}_{i^{*}})$ that aligns the target item $i^{*}$ with this 
estimated direction, while a constraint-based safety gate preserves stealthiness, 
semantic validity, image-text consistency, and platform-plausible presentation. 
Empirically, \ATK\ has three properties: (i) on VIP5 it wins every (dataset, prompt) 
cell at ER@20 on MicroLens and Beauty against fifteen baselines, by 
$1.4$--$1.8\times$; (ii) the same content transfers unchanged to 
Qwen2-VL-2B-Instruct, winning all three zero-shot ER@20 comparisons; 
(iii) \ATK\ survives three content-filter defences ($\Delta\mathrm{ER}\!<\!0.004$) 
without degrading benign-user utility.

\paragraph{Contributions.}

\begin{itemize}[itemsep=-3pt, topsep=0pt]
    \item We identify and empirically validate a compression-to-consensus 
bias in candidate-aware MLLM-RecSys: ranking is disproportionately 
driven by cues that survive compression and receive cross-modal 
consensus. To our knowledge, we are the first to formalise criteria 
hijacking for MLLM-RecSys under a strict black-box, zero-query, 
single-seller threat model.
    \item We propose \textbf{\ATK}, a zero-query pipeline that realises 
this attack through public-encoder-guided target estimation and 
stealthiness-constrained content construction.
    \item Experiments across three datasets and two victim architectures 
demonstrate substantial target-exposure gains while preserving benign-user utility.
\end{itemize}

\section{Related Work}
\label{sec:related-work}

\paragraph{Multimodal LLM-based Recommender Systems.}
Recent advances in MLLMs have reshaped recommender systems from 
ID-centric matching toward content-grounded architectures by 
integrating textual descriptions, visual content, and user 
behaviour~\citep{geng2023vip5,ye2025harnessing,giahi2025vl}. 
Representative systems include VIP5~\citep{geng2023vip5}, which 
unifies vision, language, and personalisation under a 
candidate-aware MLLM framework via parameter-efficient fine-tuning, 
and MLLM-MSR~\citep{ye2025harnessing}, which compares candidates 
against compressed multimodal preference summaries. Despite their 
growing deployment, the security of MLLM-RecSys under cross-modal 
compression and fusion remains under-investigated.

\paragraph{Adversarial Attacks on Recommender Systems.}
Existing attacks span four categories. \emph{Interaction-level 
attacks} manipulate user-item records (e.g., 
RandAttack~\citep{o2005recommender}, 
PopAttack~\citep{zhou2015shilling}) but yield limited targeted 
influence on content-driven MLLM-RecSys, where ranking is 
governed by fused multimodal 
representations~\citep{guan2026venomrec}. \emph{Single-modality 
content perturbations} edit either 
text~\citep{jin2020bert,li2020bert,gao2018black,li2018textbugger,ribeiro2020beyond} 
or 
image~\citep{liu2021adversarial,chen2024adversarial,moosavi2017universal,zhou2023advclip,zhang2026dualtap} 
at inference time, but cross-modal consensus often suppresses 
such attacks: the unperturbed modality acts as a semantic anchor 
that rectifies inconsistency. \emph{Cross-modal training-time 
poisoning} attacks (VenomRec~\citep{guan2026venomrec}, 
Shadowcast~\citep{xu2024shadowcast}) address this resilience by 
synchronising perturbations across modalities but require 
access to fine-tuning data or training pipelines. 
\emph{Query-based attacks} such as 
CheatAgent~\citep{ning2024cheatagent} and stealthy LLM-RecSys 
methods~\citep{zhang2024stealthy} estimate adversarial 
directions from victim-output feedback. Both regimes assume 
capabilities forbidden under strict zero-query inference-time 
deployment.

\paragraph{Inference-time Decision Criteria Hijacking.}
A complementary line of work studies criteria hijacking against 
text rankers~\citep{qian2025ranking}, where adversaries 
exploit how ranking models score relevance rather than force 
misclassification. This perspective informs our threat model but 
applies to text-only ranking; the compression-to-consensus 
mechanism we exploit is unique to candidate-aware multimodal 
RecSys. Black-box transfer attacks~\citep{papernot2017practical} 
on image classifiers similarly use surrogate models, but lack a structural mechanism for cross-modal 
coordination. To our knowledge, \ATK\ is the first attack 
against MLLM-RecSys to combine all four properties: strict 
zero-query operation, inference-time content editing, 
single-seller scope, and cross-modal coordination grounded in a 
structural compression-to-consensus bias.

\section{Problem Formulation}
\label{sec:formulation}

\subsection{Victim MLLM-RecSys Framework}
\label{subsec:victim}

We consider candidate-aware MLLM-RecSys, exemplified by 
VIP5~\citep{geng2023vip5} and MLLM-MSR~\citep{ye2025harnessing}, 
which cast recommendation as sequence-to-sequence generation over 
compressed multimodal item representations. Let $\mathcal{U}$ and 
$\mathcal{I}$ denote the user and item sets; each item 
$i \in \mathcal{I}$ carries a textual title/description $t_i$ and 
a visual product image $v_i$.

\paragraph{Prompt construction and ranking.}
For a user $u$ with interaction history 
$\mathcal{H}_u = [i_1, \dots, i_L]$, each item $i_k$ is encoded 
into a multimodal token sequence 
$\mathbf{e}_{k} = \tau(t_{i_k}) \oplus \zeta(g(v_{i_k}))$, where 
$\tau(\cdot)$ is the text tokeniser, $g(\cdot)$ is a frozen visual 
encoder (\eg CLIP-ViT~\citep{radford2021learning}), and 
$\zeta(\cdot)$ is a trainable projector. The user-side prompt is constructed as
\begin{equation}
\label{eq:prompt}
    X_u = \mathcal{P} \oplus \tau(u) \oplus [\mathbf{e}_{1}, \dots, \mathbf{e}_{L}],
\end{equation}
with $\mathcal{P}$ an instruction template and $\tau(u)$ a 
textualised user identifier. The MLLM's attention layers serve as 
the cross-modal fusion mechanism $\Phi_{\Theta^{*}}$. 
At serving time, the deployed recommender $f_{\Theta^{*}}$
scores each candidate $i \in \mathcal{C}_u$ using user
context and the candidate's multimodal content $(t_i,v_i)$,
and returns the top-$K$ items.
$\Theta^{*}$ denotes the victim-specific parameters separating 
$f_{\Theta^{*}}$ from any public backbones (\eg the lightweight 
adapters in PEFT-style instantiations like VIP5). Because scoring 
runs through the fused representation, any edit to $(t_i, v_i)$ 
propagates through $\Phi_{\Theta^{*}}$ and can shift the item's 
rank. We focus on the ranking stage with a given candidate set 
$\mathcal{C}_u \subseteq \mathcal{I}$ containing the target 
$i^{*}$.

\subsection{Decision Criteria Hijacking}
\label{subsec:threat}

We study targeted promotion against a deployed MLLM recommender 
$f_{\Theta^{*}}$ with frozen parameters $\Theta^{*}$. The 
adversary is a seller who owns a single target item 
$i^{*} \in \mathcal{I}$ and wants to raise its exposure by editing 
its published text and image, a standard capability on major 
e-commerce platforms. Adapting the framework of~\citet{qian2025ranking} from text rankers to candidate-aware multimodal RecSys, we formalise this inference-time attack as decision criteria hijacking.

\begin{definition}[Decision Criteria Hijacking]
\label{def:hijack}
Given a deployed victim recommender $f_{\Theta^{*}}$ with frozen 
parameters $\Theta^{*}$, an attack is a tuple 
$\mathcal{A} = (i^{*},\, T_\phi,\, \mathcal{B})$, where:
\begin{itemize}[itemsep=1pt, topsep=2pt, leftmargin=1.5em]
    \item $i^{*} \in \mathcal{I}$ is the single target item owned by the seller;
    \item $T_\phi : (t_{i^{*}},\, v_{i^{*}}) \mapsto (\tilde{t}_{i^{*}},\, \tilde{v}_{i^{*}})$ maps the original textual content and image to an edited pair;
    \item $\mathcal{B}$ is a stealthiness constraint set bounding the edited content.
\end{itemize}
\end{definition}
The attack hijacks $f_{\Theta^{*}}$'s \emph{decision criteria}, 
the structural preferences it uses to score candidates given their content, by shaping $i^{*}$'s content via $T_\phi$ to align with them rather than by deceiving the model.

\paragraph{Stealthiness Constraints $\mathcal{B}$.}
The constraint set $\mathcal{B}$ requires 
$(\tilde{t}_{i^{*}}, \tilde{v}_{i^{*}})$ to satisfy: 
(i) \emph{visual fidelity}, $\|\tilde{v}_{i^{*}} - v_{i^{*}}\|_p \le \varepsilon_v$; 
(ii) \emph{category preservation}, 
$\mathrm{sim}_{\mathrm{cat}}(\tilde{v}_{i^{*}},\, c_{i^{*}}) \ge \tau_{\mathrm{cat}}$; 
(iii) \emph{text-content fidelity}, 
$\mathrm{sim}_{\mathrm{text}}(\tilde{t}_{i^{*}},\, t_{i^{*}}) \ge \tau_{\mathrm{orig}}$; 
and (iv) \emph{cross-modal coherence}, 
$\mathrm{sim}_{\mathrm{CLIP}}(\tilde{t}_{i^{*}},\, \tilde{v}_{i^{*}}) \ge \tau_{\mathrm{align}}$. 
Conditions (i)--(iii) enforce unimodal naturalness; (iv) enforces 
cross-modal coherence.

\paragraph{Attacker's Knowledge and Capabilities.}
We assume a strict black-box, zero-query setting. The adversary 
has no access to $\Theta^{*}$, gradients, fine-tuning data, or any 
query interface to $f_{\Theta^{*}}$. Only $i^{*}$ may be modified; 
all other items, users, and training data are untouched. Within 
$i^{*}$, the adversary may propose any $T_\phi$ satisfying 
$\mathcal{B}$, and may use publicly available pretrained encoders 
(\eg CLIP) on its own device to plan the edit. These encoders are 
never calibrated against $f_{\Theta^{*}}$; their overlap with 
frozen sub-components of the victim pipeline is a public-knowledge 
prior, not query access.

\paragraph{Attacker's Goal.}
The adversary aims to maximise the fraction of users for whom 
$i^{*}$ appears in the top-$K$ list. We define the \emph{Exposure 
Rate} as
\begin{equation}
\label{eq:er}
    \mathrm{ER}_{K}(i^{*};\, \tilde{t}_{i^{*}},\, \tilde{v}_{i^{*}})
    \;=\; \frac{1}{|\mathcal{U}|} \sum_{u \in \mathcal{U}}
    \mathbb{I}\!\Big[\, i^{*} \in \mathrm{Top}\text{-}K\!\big(u;\, f_{\Theta^{*}},\, \tilde{t}_{i^{*}},\, \tilde{v}_{i^{*}}\big) \,\Big],
\end{equation}
where only $i^{*}$'s content is replaced by 
$(\tilde{t}_{i^{*}}, \tilde{v}_{i^{*}})$. The attack reduces to
\begin{equation}
\label{eq:opt}
    \max_{T_\phi}\;\; \mathrm{ER}_{K}\!\big(i^{*};\; T_\phi(t_{i^{*}},\, v_{i^{*}})\big)
    \quad \mathrm{s.t.} \quad
    T_\phi(t_{i^{*}},\, v_{i^{*}}) \in \mathcal{B}(t_{i^{*}},\, v_{i^{*}}).
\end{equation}
We evaluate on an attackable subset 
$\mathcal{I}^{\dagger} \subseteq \mathcal{I}$ of items that the 
clean recommender does not already place prominently, so the 
measured effect is not confounded with the recommender's existing 
bias. This target selection is performed solely for evaluation;
\ATK\ does not use clean-victim predictions during attack
construction. HR@$K$ and NDCG@$K$ on benign users are tracked as a utility 
check.

\section{Empirical Evidence for the Compression-to-Consensus Bias}
\label{sec:mechanism}

We validate two claims on the VIP5 victim:
(i) C2C-oriented content manipulation achieves substantial
target-exposure gains even with random hero selection from
the pre-filtered pool, and (ii) joint scoring
$S(f)=\tilde{\kappa}(f)\!\cdot\!\tilde{\chi}(f)$
further improves exposure over random and single-axis selection.

\paragraph{Setup.}

Holding modules M1, M5--M7 fixed as in the full \ATK\
pipeline (Alg.~\ref{alg:c3h}), we vary only the M4
hero-selection criterion across four conditions:
\emph{Random} (uniform draw from $\mathcal{F}$), 
\emph{$\tilde{\kappa}$-only}, \emph{$\tilde{\chi}$-only}, and 
\emph{Full C2C}. All four use only public CLIP encoders (B/16 for 
$V$, B/32 for $A_{\text{vis}}$) and issue no victim query. Targets 
are the $N{=}9$ MicroLens attackable items, replicated on $N{=}14$ 
Beauty targets; ER@20 is averaged over 1k benign users per cell.

\begin{table}[h]
  \centering
  \small
  \caption{Cue-selection ablation on \textbf{MicroLens} ($N{=}9$). 
ER@20 over 1k benign users per cell.}
  \label{tab:mechanism}
  \setlength{\tabcolsep}{6pt}
  \renewcommand{\arraystretch}{1.20}
  \begin{tabular}{l | c c c c >{\columncolor{blue!8}}c}
    \toprule
    \rowcolor{yellow!15}
    \textbf{M4 criterion}
      & \textbf{NoAttack}
      & \textbf{Random}
      & \textbf{$\tilde{\kappa}$-only}
      & \textbf{$\tilde{\chi}$-only}
      & \textbf{\ATK\ (Full C2C)} \\
    \midrule
    \textbf{ER@20} \,(Few-Shot)
      & 0.0000
      & 0.1732
      & 0.1752
      & 0.1749
      & \underline{\textbf{0.2727}} \\
    \textbf{ER@20} \,(Zero-Shot)
      & 0.0000
      & 0.1761
      & 0.1787
      & 0.1783
      & \underline{\textbf{0.2739}} \\
    \bottomrule
  \end{tabular}
\end{table}

\paragraph{Evidence consistent with the C2C bias (Tab.~\ref{tab:mechanism}).}
Random hero selection within $\mathcal{F}$ already attains ER@20 
of $0.173$ (FS) / $0.176$ (ZS), well above the NoAttack floor. 
Since Random introduces no cue-prioritisation signal, this gain
is observed without explicit cue ranking; M1 pre-filters the
pool by cross-modal CLIP similarity. Single-axis $\tilde{\kappa}$- or 
$\tilde{\chi}$-only selection stays within $\pm 0.003$ of Random. 
These observations indicate limited gains from selecting on
either axis alone within the pre-filtered pool, motivating
the joint scoring strategy.

\paragraph{Joint scoring further refines selection within the pool.}
Full C2C scoring lifts average ER@20 by approximately
$+0.099$ (FS) / $+0.098$ (ZS) over Random, reaching
$0.273$ / $0.274$ (Tab.~\ref{tab:mechanism}).
These results support joint scoring as an empirically motivated cue-selection heuristic in these settings.

\paragraph{Interpretation of the two axes.}
The scores $\tilde{\kappa}$ and $\tilde{\chi}$ represent
compression survivability and cross-modal consensus,
respectively. In the cue-selection ablation
(Tab.~\ref{tab:mechanism}), Full C2C achieves higher
average ER@20 than either single-axis selection criterion.
This observation supports joint scoring as an empirical heuristic for cue selection in the evaluated settings, but does not establish statistical independence between the scores or a causal mechanism in the victim.

\section{Method: \ATK}
\label{sec:method}

\subsection{Method Overview}
\label{sec:method-overview}

\paragraph{Surrogate decomposition.}
Eq.~(\ref{eq:opt}) cannot be solved directly under the zero-query 
setting: its outer objective $\mathrm{ER}_{K}$ depends on the 
victim $f_{\Theta^{*}}$, whose parameters and per-user predictions 
are hidden. Following the surrogate-transfer pattern for black-box 
attacks~\citep{papernot2017practical}, we replace $\mathrm{ER}_K$ 
with a surrogate $\hat{\phi}(\tilde{t}, \tilde{v})$ that scores 
content by alignment with the C2C bias using only public encoders. 
We hypothesise that the input-side compression structure is driven by the public, frozen backbones (CLIP-ViT, T5) rather 
than by the lightweight fine-tuned adapters 
$\Theta^{*}$~\citep{radford2021learning,achille2018emergence}, so 
cue rankings produced on public encoders should transfer to the 
victim, which we test empirically in Sec.~\ref{sec:mechanism}.

\begin{figure*}[h]
    \centering
    \includegraphics[width=1\linewidth]{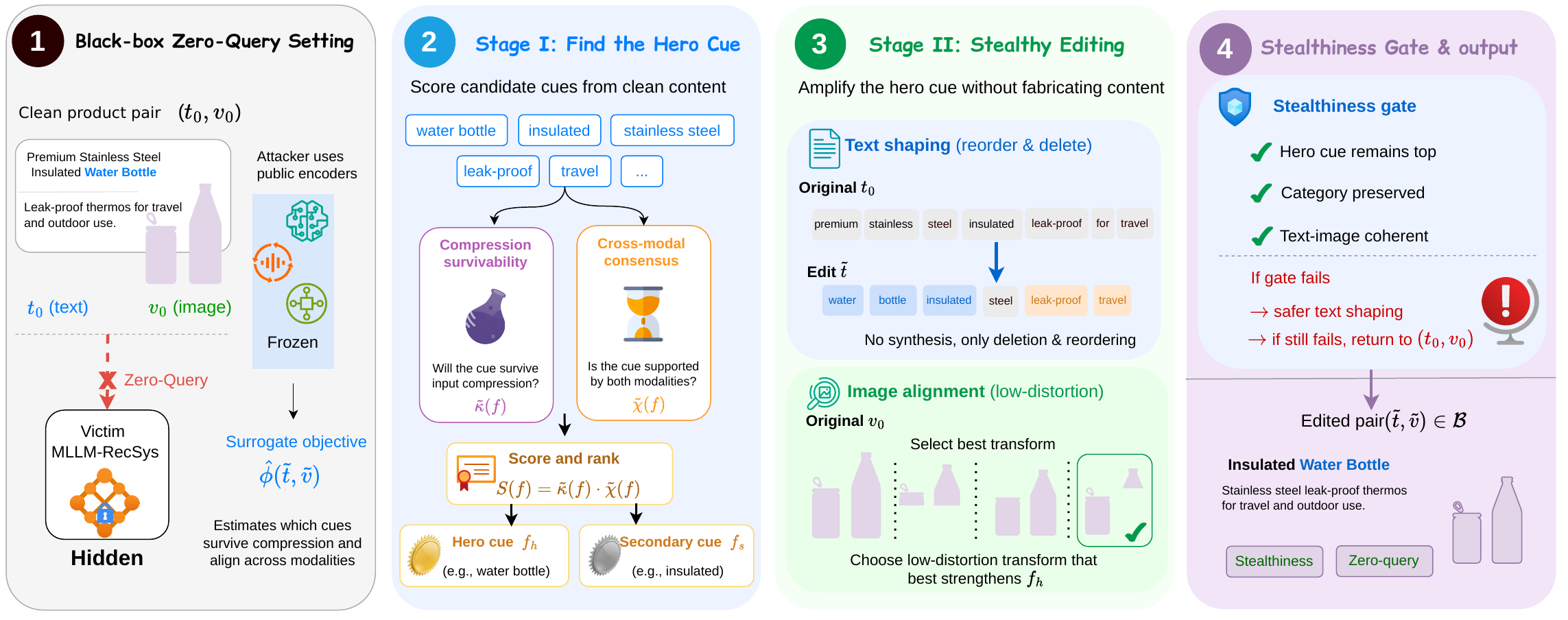}
    \caption{The schematic overview of \ATK. Under a strict zero-query setting, the framework operates in two stages: (1) \textbf{Cue Scoring (Stage~I)} ranks candidate cues from $(t_0, v_0)$ by compression-survivability $\hat{\kappa}$ and cross-modal consensus $\hat{\chi}$, selecting a hero cue $f_h$ and a secondary cue $f_s$ by joint score $S=\tilde{\kappa}\cdot\tilde{\chi}$. (2) \textbf{Stealthy Editing (Stage~II)} amplifies $f_h, f_s$ via text shaping and a low-distortion image transform, gated against the stealthiness set $\mathcal{B}$.}
    \label{fig:placeholder}
\end{figure*}

\paragraph{Two-stage framework.}
\ATK\ realises this surrogate optimisation as a two-stage pipeline 
(Fig.~\ref{fig:placeholder}). 
For brevity we abbreviate $(t_{i^{*}}, v_{i^{*}})$ as $(t_{0}, v_{0})$ 
and $(\tilde{t}_{i^{*}}, \tilde{v}_{i^{*}})$ as 
$(\tilde{t}, \tilde{v})$. Stage~I scores cues along 
compression-survivability and cross-modal consensus from the clean 
content alone, returning a hero cue $f_{h}$ and a secondary cue 
$f_{s}$. Stage~II builds $(\tilde{t}, \tilde{v})$ by shaping the text first, 
then aligning the image conditioned on the new text, and finally 
checking stealthiness with a clean-baseline fallback. The pipeline is deterministic, runs once per target, and 
issues no queries to $f_{\Theta^{*}}$.

\subsection{Stage I: Cue Scoring and Hero Selection}
\label{sec:stage1}

\paragraph{Cue mining.}
We extract a candidate cue set $\mathcal{F}$ from $(t_{0}, v_{0})$ 
by collecting noun-phrase chunks of $t_{0}$ and keeping only spans 
whose CLIP cross-modal similarity to $v_{0}$ exceeds a low-recall 
threshold~\citep{mihalcea2004textrank,radford2021learning}. Each 
$f \in \mathcal{F}$ is annotated with a position class (title, 
body, implicit), used as a position weight for $A_{\text{text}}$ in 
the consensus score.

\paragraph{Compression-survivability scoring ($\hat{\kappa}$).}
A cue contributes to ranking only if its signal survives the 
victim's input compression. Information-bottleneck theory predicts 
that compression preserves task-relevant 
features~\citep{tishby2000information,achille2018emergence}; 
$\hat{\kappa}$ takes an OR-gate form measuring whether a cue 
remains visible on at least one modality:
\begin{equation}
\hat{\kappa}(f) = 1 - \bigl(1 - C(f)\bigr)\bigl(1 - V(f)\bigr),
\label{eq:kappa}
\end{equation}
where $C(f)$ is the TextRank centrality of $f$ in $t_{0}$ averaged 
over text truncations of length $L_{\text{text}} \in \{20, 40, 80\}$, 
and $V(f)$ is the mean attention weight on the 
top-$\lceil P/4 \rceil$ patches matching $f$ (by cosine similarity 
to the cue text embedding) under CLIP-ViT-B/16 attention 
rollout~\citep{abnar2020quantifying}, with $P{=}196$ at $224{\times}224$ input image resolution.

\paragraph{Cross-modal consensus scoring ($\hat{\chi}$).}
Reaching fusion is necessary but not sufficient: pure 
attention layers tend toward rank collapse, so cues without 
cross-modal support fail to amplify during 
fusion~\citep{dong2021attention}. $\hat{\chi}$ therefore requires 
both modalities to support the cue (AND-gate):
\begin{equation}
\hat{\chi}(f) = A_{\text{text}}(f) \cdot A_{\text{vis}}(f),
\label{eq:chi}
\end{equation}

where $A_{\text{text}}(f) \in \{1.0, 0.5, 0.2\}$ denotes the predefined position weight assigned to each cue (title/body/implicit), and $A_{\text{vis}}(f)$ is computed as the mean cosine similarity between the original image $v_{0}$ and prompt templates of the form ``a photo of \{$f$\}'' under CLIP-ViT-B/32.

\paragraph{Aggregation and hero selection.}
We rank-normalise each axis as 
$\tilde{\kappa}(f) = 1 - (r_{\kappa}(f) - 1)/|\mathcal{F}|$ 
($\tilde{\chi}$ analogously), with $r_\kappa, r_\chi$ the 
descending ranks on $\hat{\kappa}, \hat{\chi}$. The 
compression-to-consensus score is 
$S(f) = \tilde{\kappa}(f) \cdot \tilde{\chi}(f)$, and the hero and 
secondary cues are 
$(f_{h}, f_{s}) = \mathrm{top}\text{-}2_{f \in \mathcal{F}}\, S(f)$, 
with ties at $|S(f_{h}) - S(f_{s})| < 0.05$ broken by the 
higher-$\hat{\chi}$ cue.\footnote{When $|\mathcal{F}| < 4$, we use 
$\hat{\kappa} \cdot \hat{\chi}$ directly. The CLIP-based components 
of $\hat{\kappa}$ and $\hat{\chi}$ use disjoint backbones (B/16 
patch-level, B/32 global), so the two axes draw on architecturally 
independent signals.}

\subsection{Stage II: Constructing Content within $\mathcal{B}$}
\label{sec:stage2}

Stage~II amplifies $f_{h}$ in the edited content while keeping the 
result inside $\mathcal{B}$. We define the surrogate objective 
$\hat{\phi}(\tilde{t}, \tilde{v}) = S(f_{h}; \tilde{t}, \tilde{v})$ 
as the C2C score of $f_{h}$ recomputed via 
Eqs.~\eqref{eq:kappa}--\eqref{eq:chi} on edits: larger 
$\hat{\phi}$ means $f_h$'s signal survives compression more strongly 
and aligns more tightly across modalities. Maximising $\hat{\phi}$ 
gives a query-free proxy for the un-evaluable $\mathrm{ER}_{K}$ 
(validated empirically in Sec.~\ref{sec:experiments}).

\paragraph{Text shaping.}
The text-shaping operator amplifies $f_{h}$ and $f_{s}$ by deletion 
and reordering of $t_{0}$, with no token synthesis. It marks stop 
words, fillers, and negative-sentiment tokens for deletion, applies 
a white-list override to keep tokens overlapping $f_{h}$ or $f_{s}$, 
and reorders surviving tokens with $f_{h}$ tokens first, $f_{s}$ 
next, other cue tokens by descending $S(f)$, and leftover content 
tokens last. This guarantees $\tilde{t} \subseteq t_{0}$ as a 
multiset (no fabricated content); $\tau_{\text{orig}}$ is then 
checked by the stealthiness gate.

\paragraph{Image alignment.}
The image-aligning operator picks the best transform from a finite 
family $\mathcal{T} = \{\theta_{0}, \dots, \theta_{4}\}$ that 
covers the low-distortion edits typical of e-commerce product 
photography (spatial framing and photometric adjustment) and 
excludes geometric distortions that would visibly violate listing 
conventions. The five transforms are: $\theta_{0}$, identity; 
$\theta_{1}, \theta_{2}$, saliency-guided crops centred on 
$f_{h}, f_{s}$ via Grad-CAM~\citep{selvaraju2017grad}; $\theta_{3}$, 
a $0.7\times$ centre crop; $\theta_{4}$, mild contrast and 
brightness adjustment on the hero region. With 
$\lambda = \hat{\chi}(f_{s})/\hat{\chi}(f_{h})$ (aggregation aborts 
if $\hat{\chi}(f_{h}) = 0$), the selection rule scores each 
transform by
\begin{equation}
\mathrm{score}(\theta) = S(f_{h};\, \tilde{t}, T_{\theta}(v_{0})) 
+ \lambda \cdot S(f_{s};\, \tilde{t}, T_{\theta}(v_{0})),
\label{eq:m6-score}
\end{equation}
and sets $\tilde{v} = T_{\theta^{*}}(v_{0})$ with 
$\theta^{*} = \arg\max_{\theta \in \mathcal{T}} \mathrm{score}(\theta)$. 
Both $\hat{\kappa}$ and $\hat{\chi}$ are recomputed per candidate, capturing position shifts from text reordering. Visual 
fidelity follows from the family design 
itself.\footnote{For image side lengths below $200$~px, $\mathcal{T}$ falls 
back to $\{\theta_{0}, \theta_{4}\}$; below $100$~px, to 
$\{\theta_{0}\}$.}

\paragraph{Stealthiness gate.}
The gate checks $(\tilde{t}, \tilde{v}) \in \mathcal{B}$ in four 
sequential phases: (1) re-check the hero-cue argmax under 
recomputed $\hat{\kappa}, \hat{\chi}$; (2) the 
category-preservation gate at $\tau_{\text{cat}}=0.20$; (3) the 
text-fidelity gate ($\tau_{\text{orig}}=0.18$) and cross-modal 
coherence gate ($\tau_{\text{align}}=0.18$) under 
CLIP-ViT-B/32~\citep{radford2021learning}; on Phase~1--3 failure, 
(4) re-run text shaping with safest-prefix-only deletions and 
re-test. If Phase~4 fails, the gate uses $(t_{0}, v_{0})$. Pseudocode and hyperparameters are in Alg.~\ref{alg:c3h}
and App.~\ref{app:cross-arch}, respectively.

\paragraph{Pipeline Pseudocode.}
Alg.~\ref{alg:c3h} summarises the full \ATK\ pipeline described above.

\begin{algorithm}[h]
\caption{\ATK\ Pipeline. Stage~I (M1--M4) scores cues for the C2C bias via public encoders; Stage~II (M5--M7) constructs $(\tilde{t}, \tilde{v})$ within the stealthiness set $\mathcal{B}$.}
\label{alg:c3h}
\begin{algorithmic}[1]
\Require Clean content $(t_{0}, v_{0})$, target category $c_{i^{*}}$, public encoders $E_{16}$ (B/16) and $E_{32}$ (B/32), thresholds $\tau_{\text{cat}}, \tau_{\text{orig}}, \tau_{\text{align}}$, transform family $\mathcal{T} = \{\theta_{0}, \dots, \theta_{4}\}$
\Ensure Edited pair $(\tilde{t}, \tilde{v}) \in \mathcal{B}$, or clean baseline $(t_{0}, v_{0})$
\State $\mathcal{F} \gets \textsc{CueMine}(t_{0}, v_{0})$ \Comment{M1}
\If{$|\mathcal{F}| < 2$}
    \State \Return $(t_{0}, v_{0})$ \Comment{insufficient cues}
\EndIf
\For{$f \in \mathcal{F}$}
    \State $\hat{\kappa}(f) \gets 1 - (1 - C(f))(1 - V(f; E_{16}))$ \Comment{M2, Eq.~\eqref{eq:kappa}}
    \State $\hat{\chi}(f) \gets A_{\text{text}}(f) \cdot A_{\text{vis}}(f; E_{32})$ \Comment{M3, Eq.~\eqref{eq:chi}}
\EndFor
\If{$|\mathcal{F}| < 4$}
    \State $S(f) \gets \hat{\kappa}(f) \cdot \hat{\chi}(f)$ \Comment{M4}
\Else
    \State $\tilde{\kappa}, \tilde{\chi} \gets \textsc{RankNorm}(\hat{\kappa}, \hat{\chi})$
    \State $S(f) \gets \tilde{\kappa}(f) \cdot \tilde{\chi}(f)$
\EndIf
\State $(f_{h}, f_{s}) \gets \mathrm{top}\text{-}2_{f \in \mathcal{F}} \, S(f)$ \Comment{tiebreak by $\hat{\chi}$ when $|S(f_{h})-S(f_{s})| < 0.05$}
\If{$\hat{\chi}(f_{h}) = 0$}
    \State \Return $(t_{0}, v_{0})$ \Comment{abort aggregation}
\EndIf
\State $\tilde{t} \gets \textsc{ShapeText}(t_{0}, f_{h}, f_{s})$ \Comment{M5: delete + reorder}
\State $\lambda \gets \hat{\chi}(f_{s}) / \hat{\chi}(f_{h})$
\For{$\theta \in \mathcal{T}$}
    \State $\mathrm{score}(\theta) \gets S(f_{h};\, \tilde{t}, T_{\theta}(v_{0})) + \lambda \cdot S(f_{s};\, \tilde{t}, T_{\theta}(v_{0}))$ \Comment{M6, Eq.~\eqref{eq:m6-score}}
\EndFor
\State $\theta^{*} \gets \arg\max_{\theta} \mathrm{score}(\theta)$; \; $\tilde{v} \gets T_{\theta^{*}}(v_{0})$
\If{Phase~1--3 of M7 pass on $(\tilde{t}, \tilde{v})$} \Comment{M7 verification}
    \State \Return $(\tilde{t}, \tilde{v})$
\ElsIf{Phase~4 yields $(\tilde{t}_{\text{fb}}, v_{0})$ passing Phases~1--3}
    \State \Return $(\tilde{t}_{\text{fb}}, v_{0})$
\Else
    \State \Return $(t_{0}, v_{0})$ \Comment{abort to clean baseline}
\EndIf
\end{algorithmic}
\end{algorithm}

\section{Experimental Evaluation}
\label{sec:experiments}

\subsection{Experimental Setup}
\label{sec:setup}

\paragraph{Datasets and Victim Models.}
We evaluate on three multimodal recommendation benchmarks:
\textbf{MicroLens-100K}~\citep{ni2025content}, a Chinese short-video
corpus, and \textbf{Amazon Beauty} and \textbf{Amazon
Sports}~\citep{mcauley2013hidden}. Following prior
work~\citep{geng2023vip5,zhang2024stealthy}, we use pre-extracted
CLIP-ViT-B/32 visual embeddings~\citep{radford2021learning} and
leave-one-out evaluation. The primary victim is
\textbf{VIP5}~\citep{geng2023vip5}, a candidate-aware MLLM-RecSys
built on a frozen T5-small backbone~\citep{raffel2020exploring} and
fine-tuned via PEFT adapters~\citep{sung2022vl}. For
cross-architecture transfer we additionally use
\textbf{Qwen2-VL-2B-Instruct}~\citep{wang2024qwen2} under few-shot and
zero-shot prompts.

\paragraph{Baselines.}
We compare \ATK\ against fifteen representative attack baselines
across three categories.
\textbf{Interaction-level attacks}: RandAttack~\citep{o2005recommender}
and PopAttack~\citep{zhou2015shilling}.
\textbf{Textual attacks}: TextFooler~\citep{jin2020bert},
DeepWordBug~\citep{gao2018black}, BERT-Attack~\citep{li2020bert},
TextBugger~\citep{li2018textbugger}, CheckList~\citep{ribeiro2020beyond},
and CheatAgent~\citep{ning2024cheatagent}.
\textbf{Image attacks}: EXPA, INSA,
SEMA~\citep{liu2021adversarial}, VAA~\citep{cohen2021black},
IPDGI~\citep{chen2024adversarial}, UAP~\citep{moosavi2017universal},
and AdvCLIP~\citep{zhou2023advclip}.

\paragraph{Evaluation Metrics.}
\textbf{Effectiveness}: Exposure Rate (ER@$K$), the fraction of
users for whom $i^{*}$ appears in the top-$K$ list.
\textbf{Stealthiness}: ROUGE-$1/2/L$~\citep{lin2004rouge} for
textual fidelity, and the semantic-alignment gap SA~$\Delta$
(absolute change in CLIP-ViT-B/32 image-text cosine similarity
relative to the clean item) for cross-modal coherence.
\textbf{Utility}: HR@$K$ and
NDCG@$K$~\citep{jarvelin2002cumulated} on benign users. We use
$K \in \{10, 20\}$ throughout.

\paragraph{Implementation Details.}
We implement all models in PyTorch~1.13.1 (CUDA~11.7) and fine-tune VIP5 
via AdamW~\citep{loshchilov2017decoupled}. The VIP5 victim is trained 
independently for each of MicroLens, Beauty, and Sports for 20 epochs 
at learning rate $1\!\times\!10^{-3}$, effective batch size 128, 
warmup ratio 0.1, and gradient clipping at 5.0; each fine-tuning run 
takes approximately 10 hours on 4$\times$ NVIDIA L20 GPUs. The 
Qwen2-VL-2B victim is evaluated under few-shot and zero-shot prompts on the same
hardware. The \ATK\ pipeline is deterministic and runs in 
approximately 8 seconds per target when distributed across 4 GPUs 
(${\sim}34$\,s single-GPU equivalent), with no queries to the victim 
during attack construction. For each (target, prompt) setting, we fix a
single evaluation-user subset that is shared across NoAttack and all
compared attack methods: VIP5 settings score $1{,}000$ benign users each,
and Qwen2-VL settings subsample users to limit inference cost.

\subsection{Overall Performance Evaluation}
\label{sec:overall-results}

\begin{table*}[t]
  \centering
  \scriptsize
  \vspace{-2mm}
  \caption{Main results of attack effectiveness and recommendation 
utility on \textbf{MicroLens} and \textbf{Beauty} under Few-Shot 
and Zero-Shot prompts. Best ER per cell is \underline{\textbf{bolded 
and underlined}}.}
  \label{tab:main-results}
  \begin{adjustbox}{max width=\textwidth}
  \setlength{\tabcolsep}{4pt}
  \renewcommand{\arraystretch}{1.15}
  \begin{tabular}{l | cc >{\columncolor{blue!8}}c | cc >{\columncolor{blue!8}}c || cc >{\columncolor{blue!8}}c | cc >{\columncolor{blue!8}}c}
    \toprule
    \rowcolor{yellow!15}\multirow{1}{*}{} & \multicolumn{6}{c||}{\textit{Few-Shot (MicroLens)}} & \multicolumn{6}{c}{\textit{Zero-Shot (MicroLens)}} \\
    \cmidrule(lr){1-7} \cmidrule(lr){8-13}
    \multirow{2}{*}{\textbf{Attack}}     & \multicolumn{3}{c|}{\textbf{Top-10}} & \multicolumn{3}{c||}{\textbf{Top-20}}
    & \multicolumn{3}{c|}{\textbf{Top-10}} & \multicolumn{3}{c}{\textbf{Top-20}} \\
    \cmidrule(lr){2-4}\cmidrule(lr){5-7} \cmidrule(lr){8-10}\cmidrule(lr){11-13}
    & \textbf{HR} & \textbf{NDCG} & \textbf{ER} & \textbf{HR} & \textbf{NDCG} & \textbf{ER}
    & \textbf{HR} & \textbf{NDCG} & \textbf{ER} & \textbf{HR} & \textbf{NDCG} & \textbf{ER} \\
    \midrule
    NoAttack       & 0.2530 & 0.1210 & 0.0000 & 0.4320 & 0.1657 & 0.0000 & 0.2510 & 0.1227 & 0.0000 & 0.4310 & 0.1679 & 0.0000 \\
    DeepWordBug    & 0.2637 & 0.1331 & 0.0360 & 0.4458 & 0.1786 & 0.1776 & 0.2627 & 0.1349 & 0.0573 & 0.4501 & 0.1817 & 0.1854 \\
    TextFooler     & 0.2568 & 0.1263 & 0.0263 & 0.4363 & 0.1712 & 0.1671 & 0.2706 & 0.1352 & 0.0359 & 0.4523 & 0.1808 & 0.1679 \\
    BertAttack     & 0.2564 & 0.1265 & 0.0343 & 0.4391 & 0.1721 & 0.1621 & 0.2626 & 0.1318 & 0.0469 & 0.4474 & 0.1781 & 0.1523 \\
    CheatAgent     & 0.2509 & 0.1238 & 0.0319 & 0.4412 & 0.1715 & 0.1894 & 0.2574 & 0.1290 & 0.0184 & 0.4482 & 0.1768 & 0.1631 \\
    TextBugger     & 0.2620 & 0.1316 & 0.0306 & 0.4432 & 0.1770 & 0.1973 & 0.2591 & 0.1283 & 0.0381 & 0.4494 & 0.1760 & 0.1700 \\
    CheckList      & 0.2599 & 0.1295 & 0.0278 & 0.4371 & 0.1737 & 0.1734 & 0.2594 & 0.1306 & 0.0366 & 0.4398 & 0.1757 & 0.1797 \\
    RandAttack     & 0.2568 & 0.1281 & 0.0296 & 0.4428 & 0.1746 & 0.1731 & 0.2532 & 0.1261 & 0.0400 & 0.4499 & 0.1753 & 0.1679 \\
    PopAttack      & 0.2607 & 0.1282 & 0.0328 & 0.4388 & 0.1727 & 0.1673 & 0.2568 & 0.1329 & 0.0424 & 0.4414 & 0.1791 & 0.1513 \\
    EXPA           & 0.2600 & 0.1250 & 0.0326 & 0.4456 & 0.1713 & 0.1662 & 0.2654 & 0.1334 & 0.0442 & 0.4454 & 0.1784 & 0.1628 \\
    INSA           & 0.2477 & 0.1231 & 0.0231 & 0.4347 & 0.1698 & 0.1657 & 0.2631 & 0.1328 & 0.0350 & 0.4441 & 0.1780 & 0.1588 \\
    SEMA           & 0.2636 & 0.1319 & 0.0328 & 0.4504 & 0.1787 & 0.1664 & 0.2670 & 0.1331 & 0.0454 & 0.4512 & 0.1793 & 0.1638 \\
    VAA            & 0.2546 & 0.1262 & 0.0318 & 0.4356 & 0.1714 & 0.1579 & 0.2561 & 0.1284 & 0.0436 & 0.4446 & 0.1755 & 0.1569 \\
    IPDGI          & 0.2636 & 0.1339 & 0.0327 & 0.4447 & 0.1792 & 0.1593 & 0.2606 & 0.1309 & 0.0462 & 0.4484 & 0.1779 & 0.1567 \\
    UAP            & 0.2518 & 0.1256 & 0.0308 & 0.4426 & 0.1734 & 0.1748 & 0.2571 & 0.1300 & 0.0450 & 0.4471 & 0.1777 & 0.1766 \\
    AdvCLIP        & 0.2578 & 0.1267 & 0.0182 & 0.4423 & 0.1729 & 0.1852 & 0.2550 & 0.1231 & 0.0321 & 0.4446 & 0.1706 & 0.1896 \\
    \textbf{\ATK}  & 0.2554 & 0.1268 & \underline{\textbf{0.0782}} & 0.4366 & 0.1721 & \underline{\textbf{0.2727}} & 0.2488 & 0.1263 & \underline{\textbf{0.0643}} & 0.4390 & 0.1741 & \underline{\textbf{0.2739}} \\
    \hline
    \midrule
    \rowcolor{yellow!15}\multirow{1}{*}{\textbf{}} & \multicolumn{6}{c||}{\textit{Few-Shot (Beauty)}} & \multicolumn{6}{c}{\textit{Zero-Shot (Beauty)}} \\
    \cmidrule(lr){2-7} \cmidrule(lr){8-13}
    & \multicolumn{3}{c|}{\textbf{Top-10}} & \multicolumn{3}{c||}{\textbf{Top-20}}
    & \multicolumn{3}{c|}{\textbf{Top-10}} & \multicolumn{3}{c}{\textbf{Top-20}} \\
    \midrule
    NoAttack       & 0.3530 & 0.2012 & 0.0000 & 0.4870 & 0.2350 & 0.0000 & 0.3210 & 0.1829 & 0.0000 & 0.4570 & 0.2172 & 0.0000 \\
    DeepWordBug    & 0.3368 & 0.1893 & 0.0024 & 0.4808 & 0.2255 & 0.0144 & 0.3075 & 0.1681 & 0.0091 & 0.4461 & 0.2030 & 0.0306 \\
    TextFooler     & 0.3433 & 0.1935 & 0.0027 & 0.4837 & 0.2287 & 0.0166 & 0.3083 & 0.1713 & 0.0083 & 0.4456 & 0.2059 & 0.0286 \\
    BertAttack     & 0.3379 & 0.1903 & 0.0029 & 0.4768 & 0.2253 & 0.0154 & 0.3041 & 0.1673 & 0.0079 & 0.4436 & 0.2024 & 0.0288 \\
    CheatAgent     & 0.3464 & 0.1949 & 0.0024 & 0.4872 & 0.2304 & 0.0091 & 0.3113 & 0.1715 & 0.0064 & 0.4515 & 0.2068 & 0.0159 \\
    TextBugger     & 0.3370 & 0.1901 & 0.0017 & 0.4754 & 0.2249 & 0.0115 & 0.3080 & 0.1657 & 0.0089 & 0.4419 & 0.1993 & 0.0247 \\
    CheckList      & 0.3431 & 0.1947 & 0.0024 & 0.4834 & 0.2300 & 0.0116 & 0.3097 & 0.1680 & 0.0051 & 0.4453 & 0.2021 & 0.0231 \\
    RandAttack     & 0.3436 & 0.1943 & 0.0027 & 0.4885 & 0.2308 & 0.0124 & 0.3189 & 0.1737 & 0.0042 & 0.4493 & 0.2065 & 0.0167 \\
    PopAttack      & 0.3441 & 0.1954 & 0.0026 & 0.4822 & 0.2302 & 0.0149 & 0.3027 & 0.1675 & \underline{\textbf{0.0110}} & 0.4415 & 0.2024 & 0.0351 \\
    EXPA           & 0.3355 & 0.1891 & 0.0020 & 0.4808 & 0.2255 & 0.0136 & 0.3053 & 0.1662 & 0.0075 & 0.4417 & 0.2005 & 0.0277 \\
    INSA           & 0.3408 & 0.1934 & 0.0029 & 0.4836 & 0.2293 & 0.0153 & 0.3034 & 0.1650 & 0.0082 & 0.4444 & 0.2005 & 0.0303 \\
    SEMA           & 0.3442 & 0.1947 & 0.0020 & 0.4829 & 0.2297 & 0.0140 & 0.3021 & 0.1620 & 0.0082 & 0.4329 & 0.1949 & 0.0254 \\
    VAA            & 0.3404 & 0.1940 & 0.0024 & 0.4839 & 0.2300 & 0.0152 & 0.3078 & 0.1690 & 0.0071 & 0.4496 & 0.2047 & 0.0276 \\
    IPDGI          & 0.3446 & 0.1956 & 0.0024 & 0.4864 & 0.2313 & 0.0159 & 0.3103 & 0.1694 & 0.0063 & 0.4450 & 0.2033 & 0.0254 \\
    UAP            & 0.3473 & 0.1953 & 0.0023 & 0.4929 & 0.2320 & 0.0151 & 0.3103 & 0.1695 & 0.0078 & 0.4379 & 0.2016 & 0.0290 \\
    AdvCLIP        & 0.3424 & 0.1918 & 0.0027 & 0.4806 & 0.2266 & 0.0145 & 0.3034 & 0.1654 & 0.0076 & 0.4432 & 0.2006 & 0.0313 \\
    \textbf{\ATK}  & 0.3415 & 0.1917 & \underline{\textbf{0.0049}} & 0.4823 & 0.2271 & \underline{\textbf{0.0290}} & 0.3043 & 0.1657 & 0.0081 & 0.4399 & 0.1998 & \underline{\textbf{0.0623}} \\
    \bottomrule
  \end{tabular}
  \end{adjustbox}
\end{table*}

\paragraph{Attack Effectiveness.}
At $K{=}20$ on VIP5 (Tab.~\ref{tab:main-results}), \ATK\ leads in
all four (dataset, prompt) cells. On MicroLens ($N{=}9$), \ATK\
raises ER@20 to $0.273$ (FS) / $0.274$ (ZS), about $40\%$ above the
strongest baseline (TextBugger $0.197$ FS; AdvCLIP $0.190$ ZS).
Beauty is harder for content-only attacks ($N{=}14$), but \ATK\
still reaches $0.029$ (FS) / $0.062$ (ZS), close to twice the
runner-up. The near-identical MicroLens FS/ZS scores ($0.273$ vs.\
$0.274$) suggest the C2C surrogate latches onto structural decision
criteria rather than prompt-specific features. On Beauty, ZS
exceeds FS ($0.062$ vs.\ $0.029$): without exemplars to anchor the
model's ranking, criteria-aligned content faces less competition.

\paragraph{Stealthiness and Utility.}
HR@$K$ and NDCG@$K$ on benign users stay close to NoAttack across
all four cells of Tab.~\ref{tab:main-results}. Stealthiness holds
on both axes (Tab.~\ref{tab:stealth}): the edited title keeps a
high ROUGE-L with the original ($94.02$ on MicroLens, $82.22$ on
Beauty), and SA~$\Delta$ stays at $0.0020$ / $0.0135$.

\begin{table*}[t]
  \centering
  \scriptsize
  \vspace{-2mm}
  \caption{Attack stealthiness on \textbf{MicroLens} ($N{=}9$) and 
\textbf{Beauty} ($N{=}14$). ROUGE uses character-level tokenisation 
on MicroLens, word-level on Beauty.}
  \label{tab:stealth}
  \setlength{\tabcolsep}{4pt}
  \renewcommand{\arraystretch}{1.2}
  \begin{tabular}{l | ccc | >{\columncolor{blue!8}}c || ccc | >{\columncolor{blue!8}}c}
    \toprule
    \rowcolor{yellow!15}
    \multirow{1}{*}{}
        & \multicolumn{4}{c||}{\textit{MicroLens} ($N=9$)}
        & \multicolumn{4}{c}{\textit{Beauty} ($N=14$)} \\
    \cmidrule(lr){2-5}\cmidrule(lr){6-9}
    \rowcolor{yellow!15}
     & \multicolumn{3}{c|}{\textit{Unimodal Naturalness ROUGE}}
       & \textit{Cross-modal}
     & \multicolumn{3}{c|}{\textit{Unimodal Naturalness ROUGE}}
       & \textit{Cross-modal} \\
    \cmidrule(lr){2-4}\cmidrule(lr){5-5}\cmidrule(lr){6-8}\cmidrule(lr){9-9}
    \textbf{Attack}
        & \textbf{-1$\uparrow$} & \textbf{-2$\uparrow$} & \textbf{-L$\uparrow$}
        & \textbf{SA $\Delta\downarrow$}
        & \textbf{-1$\uparrow$} & \textbf{-2$\uparrow$} & \textbf{-L$\uparrow$}
        & \textbf{SA $\Delta\downarrow$} \\
    \midrule
    NoAttack
        & 100.00 & 100.00 & 100.00 & 0.0000
        & 100.00 & 100.00 & 100.00 & 0.0000 \\
    \textbf{\ATK\ }
        & 94.02 & 93.86 & 94.02 & \underline{\textbf{0.0020}}
        & 82.22 & 80.25 & 82.22 & \underline{\textbf{0.0135}} \\
    \bottomrule
  \end{tabular}
  \vspace{-2mm}
\end{table*}

\subsection{Ablation Study}
\label{sec:ablation}

We run a leave-one-out ablation (Tab.~\ref{tab:loo-ablation}) at
$K{=}20$, removing one component at a time: (1) the text operator
(image-only); (2) the image operator (text-only); and (3) the C2C
joint scoring, replaced by random hero selection from the same pool
$\mathcal{F}$.

\textbf{Dominant modality flips with the dataset.} On text-rich
MicroLens captions, removing the text operator costs more ER than
removing the image operator ($-0.110$/$-0.107$ vs.\
$-0.090$/$-0.092$ on FS/ZS). Beauty's short titles reverse the
order: dropping the image operator costs $-0.016$/$-0.048$ while
the text operator costs only $-0.009$/$-0.026$. Longer text gives
the text edit more to amplify; sparse titles shift the burden to
the visual channel.

\textbf{Joint scoring beats random hero.} Replacing
$\tilde{\kappa}\!\cdot\!\tilde{\chi}$ with a random pick from the
same pool drops ER everywhere ($-0.099$/$-0.098$ on MicroLens;
$-0.009$/$-0.025$ on Beauty), 
echoing §\ref{sec:mechanism}: pool
filtering pre-selects cross-modally aligned cues, and the joint score adds
within-pool ranking signal on top. Utility stays within $0.02$ of
full \ATK\ on every variant.

\begin{table*}[t]
  \centering
  \scriptsize
  \vspace{-2mm}
  \caption{Leave-one-out ablation on \textbf{MicroLens} ($N{=}9$) 
and \textbf{Beauty} ($N{=}14$) at $K{=}20$. $\Delta$ER is the 
drop relative to full \ATK.}
  \label{tab:loo-ablation}
  \begin{adjustbox}{max width=\textwidth}
  \setlength{\tabcolsep}{4pt}
  \renewcommand{\arraystretch}{1.2}
  \begin{tabular}{l | ccc >{\columncolor{blue!8}}c | ccc >{\columncolor{blue!8}}c || ccc >{\columncolor{blue!8}}c | ccc >{\columncolor{blue!8}}c}
    \toprule
    \rowcolor{yellow!15}
    \multirow{1}{*}{} & \multicolumn{4}{c|}{\textit{Few-Shot (MicroLens)}} & \multicolumn{4}{c||}{\textit{Zero-Shot (MicroLens)}}
                      & \multicolumn{4}{c|}{\textit{Few-Shot (Beauty)}}     & \multicolumn{4}{c}{\textit{Zero-Shot (Beauty)}} \\
    \cmidrule(lr){1-5}\cmidrule(lr){6-9}\cmidrule(lr){10-13}\cmidrule(lr){14-17}
    \textbf{Variant}
        & \textbf{HR} & \textbf{NDCG} & \textbf{ER} & \textbf{$\Delta$ER}
        & \textbf{HR} & \textbf{NDCG} & \textbf{ER} & \textbf{$\Delta$ER}
        & \textbf{HR} & \textbf{NDCG} & \textbf{ER} & \textbf{$\Delta$ER}
        & \textbf{HR} & \textbf{NDCG} & \textbf{ER} & \textbf{$\Delta$ER} \\
    \midrule
    \textbf{\ATK\ (Full C3H)} & 0.4366 & 0.1721 & \underline{\textbf{0.2727}} & --
                              & 0.4390 & 0.1741 & \underline{\textbf{0.2739}} & --
                              & 0.4823 & 0.2271 & \underline{\textbf{0.0290}} & --
                              & 0.4399 & 0.1998 & \underline{\textbf{0.0623}} & -- \\
    $-$M5 (image only)        & 0.4430 & 0.1659 & 0.1623 & $-0.110$ & 0.4488 & 0.1761 & 0.1668 & $-0.107$
                              & 0.4834 & 0.2296 & 0.0202 & $-0.009$ & 0.4500 & 0.2025 & 0.0362 & $-0.026$ \\
    $-$M6 (text only)         & 0.4430 & 0.1659 & 0.1824 & $-0.090$ & 0.4461 & 0.1746 & 0.1814 & $-0.092$
                              & 0.4747 & 0.2247 & 0.0131 & $-0.016$ & 0.4456 & 0.2035 & 0.0148 & $-0.048$ \\
    $-$C2C (random hero)      & 0.4423 & 0.1738 & 0.1732 & $-0.099$ & 0.4472 & 0.1772 & 0.1761 & $-0.098$
                              & 0.4875 & 0.2306 & 0.0196 & $-0.009$ & 0.4494 & 0.2051 & 0.0371 & $-0.025$ \\
    \midrule
    NoAttack                  & 0.4320 & 0.1657 & 0.0000 & $-0.273$ & 0.4310 & 0.1679 & 0.0000 & $-0.274$
                              & 0.4870 & 0.2350 & 0.0000 & $-0.029$ & 0.4570 & 0.2172 & 0.0000 & $-0.062$ \\
    \bottomrule
  \end{tabular}
  \end{adjustbox}
\end{table*}

\subsection{Defence Robustness}
\label{sec:defence}

We apply three content-filter defences ahead of the recommender,
each calibrated to $5$--$10\%$ rejection on clean items: AE
Filter~\citep{sakurada2014anomaly,cintas2020anomalous} (autoencoder
latent denoising in CLIP space), MCM (Maximum Concept
Matching)~\citep{ming2022delving}, and Mahalanobis OOD
detection~\citep{lee2018simple}. None meaningfully reduces \ATK\
(Tab.~\ref{tab:defense}): ER moves by less than $0.004$ under every
defence in all four cells, and HR@$K$ / NDCG@$K$ on benign users
are preserved. The reason is structural: the text operator only
deletes and reorders tokens already in $t_{0}$, and image transforms
in $\mathcal{T}$ stay within the crops and lighting adjustments
common in product photography. The edited content sits inside the
clean distribution in CLIP feature space (SA~$\Delta = 0.002$ on
MicroLens), giving outlier-based filters nothing to latch onto.

\subsection{Cross-architecture Transferability}
\label{sec:transfer}

We take the attack content $(\tilde{t}_{i^*}, \tilde{v}_{i^*})$
produced against VIP5 and apply it unchanged to
Qwen2-VL-2B-Instruct~\citep{wang2024qwen2} under few-shot and zero-shot prompts. The
two victims use different vision (Qwen's native ViT vs.\
CLIP-ViT-B/32) and language ($1.5$B Qwen-2 LM vs.\ T5-small)
backbones. Full tables are in App.~\ref{app:cross-arch}; VIP5
Sports results are in App.~\ref{app:sports}.

\ATK\ ranks first in $4$ of $6$ (dataset, prompt) cells at
$K{=}20$ on Qwen2-VL, including all three Zero-Shot cells.
Sports is the clearest case: \ATK\ wins both prompt settings
by $3.7$--$4.7\times$ over the runner-up.
The Zero-Shot $>$ Few-Shot pattern from VIP5
also reappears: at $K{=}20$, \ATK\ reaches $0.117$
(ZS) vs.\ $0.054$ (FS) on MicroLens, $0.101$ vs.\ $0.056$ on
Beauty, and $0.187$ vs.\ $0.113$ on Sports. The mechanism does not
hinge on the specific victim. On Few-Shot MicroLens and Beauty, a
few image-attack baselines edge ahead of \ATK; we attribute this
to large pixel perturbations disturbing Qwen2-VL's in-context
reasoning when few-shot exemplars are present, which is unrelated
to the C2C bias and does not generalise to Sports (the same
baselines drop below ER $0.03$ on every Few-Shot cell there).

\begin{table*}[t]
  \centering
  \scriptsize
  \vspace{-2mm}
  \caption{Defence robustness on \textbf{MicroLens} ($N{=}9$) 
under Few-Shot and Zero-Shot prompts. Each defence is 
calibrated to $5$--$10\%$ rejection on clean items.}
  \label{tab:defense}
  \begin{adjustbox}{max width=\textwidth}
  \setlength{\tabcolsep}{4pt}
  \renewcommand{\arraystretch}{1.15}
  \begin{tabular}{l | cc >{\columncolor{blue!8}}c | cc >{\columncolor{blue!8}}c || cc >{\columncolor{blue!8}}c | cc >{\columncolor{blue!8}}c}
    \toprule
    \rowcolor{yellow!15}\multirow{1}{*}{} & \multicolumn{6}{c||}{\textit{Few-Shot (MicroLens)}} & \multicolumn{6}{c}{\textit{Zero-Shot (MicroLens)}} \\
    \cmidrule(lr){1-7} \cmidrule(lr){8-13}
    \multirow{2}{*}{\textbf{Defence}}    & \multicolumn{3}{c|}{\textbf{Top-10}} & \multicolumn{3}{c||}{\textbf{Top-20}}
    & \multicolumn{3}{c|}{\textbf{Top-10}} & \multicolumn{3}{c}{\textbf{Top-20}} \\
    \cmidrule(lr){2-4}\cmidrule(lr){5-7} \cmidrule(lr){8-10}\cmidrule(lr){11-13}
    & \textbf{HR} & \textbf{NDCG} & \textbf{ER} & \textbf{HR} & \textbf{NDCG} & \textbf{ER}
    & \textbf{HR} & \textbf{NDCG} & \textbf{ER} & \textbf{HR} & \textbf{NDCG} & \textbf{ER} \\
    \midrule
    No Defence   & 0.2554 & 0.1268 & 0.0782 & 0.4366 & 0.1721 & 0.2727 & 0.2488 & 0.1263 & 0.0643 & 0.4390 & 0.1741 & 0.2739 \\
    \midrule
    AE Filter  \cite{sakurada2014anomaly,cintas2020anomalous} & 0.2571 & 0.1285 & 0.0794 & 0.4426 & 0.1748 & 0.2732 & 0.2630 & 0.1319 & 0.0638 & 0.4453 & 0.1775 & 0.2718 \\
    MCM \cite{ming2022delving}                                & 0.2604 & 0.1314 & 0.0814 & 0.4510 & 0.1793 & 0.2736 & 0.2650 & 0.1312 & 0.0628 & 0.4471 & 0.1768 & 0.2747 \\
    Mahalanobis OOD \cite{lee2018simple}                      & 0.2603 & 0.1301 & 0.0801 & 0.4479 & 0.1771 & 0.2714 & 0.2544 & 0.1278 & 0.0634 & 0.4420 & 0.1748 & 0.2709 \\
    \bottomrule
  \end{tabular}
  \end{adjustbox}
\end{table*}

\section{Conclusion}
\label{sec:conclusion}

We formalise and study decision criteria hijacking against 
MLLM-RecSys under a strict black-box, zero-query, inference-time, 
single-seller threat model. Candidate-aware MLLM-RecSys exhibit a 
compression-to-consensus (C2C) bias: the same compression and 
cross-modal fusion mechanisms that suppress conventional unimodal 
perturbations also confer disproportionate ranking influence on 
cues that survive compression and receive cross-modal 
reinforcement. \ATK\ exploits this bias as a query-free, 
public-encoder-driven pipeline that edits a single target item's 
text and image inside a stealthiness set $\mathcal{B}$.

On VIP5, \ATK\ beats fifteen baselines on every (dataset, prompt) 
cell at ER@20 by $1.1$--$1.8\times$, and the same content transfers 
unchanged to Qwen2-VL-2B-Instruct, winning all Zero-Shot cells 
across MicroLens, Beauty, and Sports, with margins reaching 
$4.7{\times}$ over the runner-up on Sports. Three anomaly-based 
content-filter defences fail to mitigate \ATK\ 
($\Delta\mathrm{ER}<0.004$), and benign-user utility (HR@$K$, 
NDCG@$K$) is preserved. Compression and cross-modal fusion are 
commonly assumed to confer robustness on MLLM-RecSys; our results 
show that they also open an on-criterion attack surface that 
requires no victim feedback.

\section{Limitations and Potential Mitigations}
\label{sec:limitations}

\ATK\ assumes edit access to both text and images, and its effectiveness may vary across targets, victim models, and platform constraints. Our evaluation covers a limited set of targets and victim models in controlled benchmarks; effectiveness on deployed platforms remains to be established. Candidate mitigations include content canonicalisation and monitoring listing edits, but their effectiveness against \ATK\ requires further evaluation, and they may reject legitimate edits or reduce recommendation utility.

\section*{Acknowledgements}
This research / project is supported by A*STAR under its Japan-Singapore Joint Call: JST-A*STAR 2024 (Project ID: R2416IR139). This research is also supported by the Ministry of Education, Singapore, under its Academic Research Fund Tier 2 (Award MOE-T2EP20125-0005). The authors declare no competing interests.

\newpage
\newpage
{
\small
\bibliographystyle{unsrtnat}
\bibliography{reference}
}

\appendix

\newpage

\section*{Appendix}

\section{Results on Sports Dataset}
\label{app:sports}
We additionally report VIP5 results on the Sports dataset 
($N{=}6$ attackable targets) for completeness. Tab.~\ref{tab:sports} 
shows that \ATK\ remains the strongest attack at $K{=}20$ on both 
prompts (FS: $0.028$ vs.\ best baseline $0.021$; ZS: $0.106$ vs.\ 
$0.096$), though the absolute margin is smaller than on MicroLens 
and Beauty due to the smaller attackable subset.
\begin{table*}[ht]
  \centering
  \scriptsize
  \caption{Main results on \textbf{Sports} ($N{=}6$) under Few-Shot 
and Zero-Shot prompts at $K{=}20$.}
  \label{tab:sports}
  \begin{adjustbox}{max width=\textwidth}
  \setlength{\tabcolsep}{4pt}
  \renewcommand{\arraystretch}{1.15}
  \begin{tabular}{l | cc >{\columncolor{blue!8}}c || cc >{\columncolor{blue!8}}c}
    \toprule
    \rowcolor{yellow!15}\multirow{1}{*}{} & \multicolumn{3}{c||}{\textit{Few-Shot (Sports)}} & \multicolumn{3}{c}{\textit{Zero-Shot (Sports)}} \\
    \cmidrule(lr){1-4} \cmidrule(lr){5-7}
    \textbf{Attack} & \textbf{HR} & \textbf{NDCG} & \textbf{ER} & \textbf{HR} & \textbf{NDCG} & \textbf{ER} \\
    \midrule
    NoAttack       & 0.3240 & 0.1666 & 0.0000 & 0.3210 & 0.1526 & 0.0000 \\
    DeepWordBug    & 0.3397 & 0.1590 & 0.0095 & 0.3373 & 0.1542 & 0.0755 \\
    TextFooler     & 0.3470 & 0.1627 & 0.0117 & 0.3305 & 0.1515 & 0.0937 \\
    BertAttack     & 0.3418 & 0.1605 & 0.0212 & 0.3290 & 0.1490 & 0.0808 \\
    CheatAgent     & 0.3385 & 0.1600 & 0.0073 & 0.3420 & 0.1534 & 0.0338 \\
    TextBugger     & 0.3385 & 0.1572 & 0.0070 & 0.3315 & 0.1541 & 0.0728 \\
    CheckList      & 0.3482 & 0.1674 & 0.0153 & 0.3338 & 0.1569 & 0.0692 \\
    RandAttack     & 0.3435 & 0.1594 & 0.0047 & 0.3293 & 0.1547 & 0.0728 \\
    PopAttack      & 0.3387 & 0.1569 & 0.0083 & 0.3282 & 0.1486 & 0.0745 \\
    EXPA           & 0.3522 & 0.1615 & 0.0177 & 0.3428 & 0.1560 & 0.0928 \\
    INSA           & 0.3420 & 0.1576 & 0.0132 & 0.3342 & 0.1527 & 0.0838 \\
    SEMA           & 0.3463 & 0.1606 & 0.0182 & 0.3348 & 0.1541 & 0.0962 \\
    VAA            & 0.3328 & 0.1539 &  0.0065 & 0.3463 & 0.1579 & 0.0618 \\
    IPDGI          & 0.3447 & 0.1604 & 0.0133 & 0.3425 & 0.1559 & 0.0745 \\
    UAP            & 0.3398 & 0.1605 & 0.0048 & 0.3400 & 0.1568 & 0.0508 \\
    AdvCLIP        & 0.3422 & 0.1572 & 0.0100 & 0.3270 & 0.1487 & 0.0702 \\
    \textbf{\ATK}  & 0.3395 & 0.1598 & \underline{\textbf{0.0282}} & 0.3380 & 0.1549 & \underline{\textbf{0.1058}} \\
    \bottomrule
  \end{tabular}
  \end{adjustbox}
\end{table*}

\section{Cross-Architecture Transferability: Results}
\label{app:cross-arch}

The cross-architecture transferability of \ATK\ on Qwen2-VL-2B-Instruct, 
summarised in §\ref{sec:transfer}, is reported in 
Tab.~\ref{tab:cross-arch} (MicroLens, $N{=}9$; Beauty, $N{=}14$) and 
Tab.~\ref{tab:cross-arch-sports} (Sports, $N{=}6$) under both 
Few-Shot and Zero-Shot prompts.
\begin{table*}[ht]
  \centering
  \scriptsize
  \caption{Cross-architecture transferability on 
\textbf{Qwen2-VL-2B-Instruct}~\citep{wang2024qwen2} for 
\textbf{MicroLens} ($N{=}9$) and \textbf{Beauty} ($N{=}14$) at $K{=}20$.}
  \label{tab:cross-arch}
  \begin{adjustbox}{max width=\textwidth}
  \setlength{\tabcolsep}{4pt}
  \renewcommand{\arraystretch}{1.15}
  \begin{tabular}{l | cc >{\columncolor{blue!8}}c | cc >{\columncolor{blue!8}}c || cc >{\columncolor{blue!8}}c | cc >{\columncolor{blue!8}}c}
    \toprule
    \rowcolor{yellow!15}\multirow{2}{*}{\textbf{Attack}} & \multicolumn{6}{c||}{\textbf{MicroLens} ($N{=}9$)} & \multicolumn{6}{c}{\textbf{Beauty} ($N{=}14$)} \\
    \cmidrule(lr){2-7} \cmidrule(lr){8-13}
    & \multicolumn{3}{c|}{\textit{Few-Shot}} & \multicolumn{3}{c||}{\textit{Zero-Shot}} & \multicolumn{3}{c|}{\textit{Few-Shot}} & \multicolumn{3}{c}{\textit{Zero-Shot}} \\
    \cmidrule(lr){2-4}\cmidrule(lr){5-7} \cmidrule(lr){8-10}\cmidrule(lr){11-13}
    & \textbf{HR} & \textbf{NDCG} & \textbf{ER} & \textbf{HR} & \textbf{NDCG} & \textbf{ER}
    & \textbf{HR} & \textbf{NDCG} & \textbf{ER} & \textbf{HR} & \textbf{NDCG} & \textbf{ER} \\
    \midrule
    NoAttack       & 0.3144 & 0.1033 & 0.0000 & 0.2534 & 0.0809 & 0.0000 & 0.3583 & 0.1563 & 0.0000 & 0.3908 & 0.1685 & 0.0000 \\
    DeepWordBug    & 0.3229 & 0.1073 & 0.0781 & 0.2398 & 0.0828 & 0.0272 & 0.3610 & 0.1491 & 0.0475 & 0.4032 & 0.1776 & 0.0194 \\
    TextFooler     & 0.3081 & 0.0978 & 0.0514 & 0.2423 & 0.0749 & 0.0223 & 0.3410 & 0.1417 & 0.0033 & 0.4149 & 0.1772 & 0.0035 \\
    BertAttack     & 0.3306 & 0.1053 & 0.0376 & 0.2742 & 0.0950 & 0.0055 & 0.3295 & 0.1361 & 0.0300 & 0.4124 & 0.1791 & 0.0053 \\
    CheatAgent     & 0.3173 & 0.1043 & 0.0533 & 0.2472 & 0.0822 & 0.0306 & 0.3558 & 0.1584 & 0.0066 & 0.4021 & 0.1830 & 0.0053 \\
    TextBugger     & 0.3193 & 0.1075 & 0.0765 & 0.2287 & 0.0819 & 0.0055 & 0.3378 & 0.1422 & 0.0532 & 0.3936 & 0.1619 & 0.0160 \\
    CheckList      & 0.3333 & 0.1060 & 0.0355 & 0.2316 & 0.0779 & 0.0327 & 0.3523 & 0.1518 & 0.0417 & 0.3929 & 0.1649 & 0.0336 \\
    RandAttack     & 0.3306 & 0.1036 & 0.0190 & 0.2310 & 0.0789 & 0.0027 & 0.3250 & 0.1461 & 0.0536 & 0.3838 & 0.1683 & 0.0018 \\
    PopAttack      & 0.3253 & 0.1057 & 0.0269 & 0.2830 & 0.0874 & 0.0220 & 0.3673 & 0.1582 & 0.0000 & 0.4142 & 0.1762 & 0.0017 \\
    EXPA           & 0.2992 & 0.0970 & 0.0943 & 0.2646 & 0.0905 & 0.0056 & 0.3289 & 0.1488 & 0.0548 & 0.4139 & 0.1685 & 0.0226 \\
    INSA           & 0.3152 & 0.1035 & 0.0870 & 0.2472 & 0.0799 & 0.0083 & 0.3356 & 0.1430 & 0.0688 & 0.4152 & 0.1732 & 0.0247 \\
    SEMA           & 0.2861 & 0.0960 & \underline{\textbf{0.0989}} & 0.2231 & 0.0735 & 0.0028 & 0.3384 & 0.1429 & 0.0643 & 0.4004 & 0.1735 & 0.0212 \\
    VAA            & 0.3431 & 0.1122 & 0.0878 & 0.2630 & 0.0831 & 0.0055 & 0.3267 & 0.1424 & 0.0783 & 0.3878 & 0.1724 & 0.0174 \\
    IPDGI          & 0.3173 & 0.1000 & 0.0907 & 0.2707 & 0.0891 & 0.0083 & 0.3261 & 0.1318 & 0.0765 & 0.3748 & 0.1597 & 0.0294 \\
    UAP            & 0.3154 & 0.1021 & 0.0755 & 0.2472 & 0.0847 & 0.0000 & 0.3570 & 0.1480 & \underline{\textbf{0.0959}} & 0.3537 & 0.1628 & 0.0470 \\
    AdvCLIP        & 0.3592 & 0.1156 & 0.0858 & 0.2598 & 0.0865 & 0.0000 & 0.3710 & 0.1666 & 0.0682 & 0.4207 & 0.1788 & 0.0414 \\
    \textbf{\ATK}  & 0.3029 & 0.0959 & 0.0536 & 0.2542 & 0.0833 & \underline{\textbf{0.1173}} & 0.3237 & 0.1420 & 0.0559 & 0.4108 & 0.1738 & \underline{\textbf{0.1014}} \\
    \bottomrule
  \end{tabular}
  \end{adjustbox}
\end{table*}

\begin{table*}[ht]
  \centering
  \scriptsize
  \caption{Cross-architecture transferability on 
\textbf{Qwen2-VL-2B-Instruct}~\citep{wang2024qwen2} for 
\textbf{Sports} ($N{=}6$) at $K{=}20$.}
  \label{tab:cross-arch-sports}
  \begin{adjustbox}{max width=\textwidth}
  \setlength{\tabcolsep}{4pt}
  \renewcommand{\arraystretch}{1.15}
  \begin{tabular}{l | cc >{\columncolor{blue!8}}c || cc >{\columncolor{blue!8}}c}
    \toprule
    \rowcolor{yellow!15}\multirow{1}{*}{} & \multicolumn{3}{c||}{\textit{Few-Shot (Sports)}} & \multicolumn{3}{c}{\textit{Zero-Shot (Sports)}} \\
    \cmidrule(lr){1-4} \cmidrule(lr){5-7}
    \textbf{Attack} & \textbf{HR} & \textbf{NDCG} & \textbf{ER} & \textbf{HR} & \textbf{NDCG} & \textbf{ER} \\
    \midrule
    NoAttack       & 0.3434 & 0.1517 & 0.0000 & 0.2877 & 0.1307 & 0.0000 \\
    DeepWordBug    & 0.3352 & 0.1481 & 0.0166 & 0.3271 & 0.1489 & 0.0141 \\
    TextFooler     & 0.3481 & 0.1576 & 0.0055 & 0.2941 & 0.1358 & 0.0118 \\
    BertAttack     & 0.3111 & 0.1536 & 0.0000 & 0.3129 & 0.1393 & 0.0000 \\
    CheatAgent     & 0.3225 & 0.1443 & 0.0000 & 0.3002 & 0.1412 & 0.0000 \\
    TextBugger     & 0.3011 & 0.1272 & 0.0304 & 0.3170 & 0.1371 & 0.0326 \\
    CheckList      & 0.2869 & 0.1266 & 0.0301 & 0.3333 & 0.1481 & 0.0399 \\
    RandAttack     & 0.2779 & 0.1180 & 0.0027 & 0.3155 & 0.1432 & 0.0093 \\
    PopAttack      & 0.3171 & 0.1325 & 0.0054 & 0.3121 & 0.1430 & 0.0000 \\
    EXPA           & 0.2877 & 0.1322 & 0.0247 & 0.3192 & 0.1380 & 0.0000 \\
    INSA           & 0.2992 & 0.1234 & 0.0216 & 0.3357 & 0.1496 & 0.0188 \\
    SEMA           & 0.3306 & 0.1506 & 0.0246 & 0.3019 & 0.1352 & 0.0024 \\
    VAA            & 0.3562 & 0.1543 & 0.0274 & 0.3341 & 0.1519 & 0.0000 \\
    IPDGI          & 0.2981 & 0.1225 & 0.0163 & 0.3341 & 0.1484 & 0.0141 \\
    UAP            & 0.3115 & 0.1245 & 0.0273 & 0.3095 & 0.1437 & 0.0000 \\
    AdvCLIP        & 0.3062 & 0.1291 & 0.0108 & 0.3066 & 0.1356 & 0.0000 \\
    \textbf{\ATK}  & 0.3113 & 0.1430 & \underline{\textbf{0.1129}} & 0.2937 & 0.1261 & \underline{\textbf{0.1865}} \\
    \bottomrule
  \end{tabular}
  \end{adjustbox}
\end{table*}

\paragraph{Hyperparameters used in our experiments.}
For all reported results, we use $\tau_{\text{cat}} = 0.20$, $\tau_{\text{orig}} = 0.18$, $\tau_{\text{align}} = 0.18$ (B/32 cosine thresholds), context window lengths $L_{\text{text}} \in \{20, 40, 80\}$ for TextRank, top-$\lceil P/4 \rceil$ patches for $V(f)$ under attention rollout, position weights $\{1.0, 0.5, 0.2\}$ for title/body/implicit in $A_{\text{text}}$, and tie-break threshold $\Delta_{\text{tie}} = 0.05$ for hero/secondary selection. Image transform parameters: $\theta_{1}, \theta_{2}$ use Grad-CAM on CLIP-ViT-B/32, with the crop bounding box expanded by $20\%$ around the cue saliency centroid; $\theta_{3}$ centre-crops to $0.7\times$ each side; $\theta_{4}$ applies contrast factor $1.15$ and brightness shift $+0.05$ within the hero region (Grad-CAM mask thresholded at the $50$th percentile). Small-image fallback: $\mathcal{T} = \{\theta_{0}, \theta_{4}\}$ when image side $< 200$~px, $\mathcal{T} = \{\theta_{0}\}$ when $< 100$~px.

\section{Compute Reporting}
\label{app:compute}

We report the computational resources used for constructing and 
evaluating \ATK. The paper reports results on three datasets 
(\textsc{MicroLens}, Amazon \textsc{Beauty}, and Amazon \textsc{Sports}); 
the compute estimates below also include additional exploratory 
experiments beyond these reported results. 
We fine-tune the VIP5 victim with parameter-efficient adapters,
generate \ATK\ content for each (dataset, target) pair using only
public encoders, and evaluate effectiveness, utility, and
stealthiness on benign user pools.
No models are trained from scratch and no external API services are 
used; all inference is performed locally.

\paragraph{Local Compute Environment.}
All experiments run on a single Ubuntu~20.04 workstation with 
$8\times$ NVIDIA L20 48~GB GPUs (CUDA 11.7), Python~3.9, and 
PyTorch~1.13.1+cu117. Source code, checkpoints, evaluation logs, 
and pre-extracted CLIP features occupy approximately $90$~GB of 
disk storage. 

\paragraph{Victim Fine-Tuning.}
We fine-tune the VIP5 victim via VL-Adapter~\cite{sung2022vl} 
(reduction factor $8$) for $20$ epochs per dataset using AdamW at 
learning rate $1\!\times\!10^{-3}$, warmup ratio $0.1$, gradient 
clipping $5.0$, an effective batch size of $128$ ($4\times 32$ across 
four L20 GPUs), and a fixed random seed of $2022$. Each dataset's 
fine-tuning takes approximately $9.5$--$10.5$ wall-clock hours; 
across the three datasets reported in the paper (\textsc{MicroLens}, 
Amazon \textsc{Beauty}, and Amazon \textsc{Sports}) and additional 
exploratory fine-tuning, victim fine-tuning consumes approximately 
$\mathbf{200}$ GPU-hours.

\paragraph{Attack Generation.}
Each \ATK\ artefact is generated once per (dataset, target) 
pair using only public CLIP-ViT-B/16 (patch-level for $\hat\kappa$) 
and CLIP-ViT-B/32 (global for $\hat\chi$) encoders. Generation is 
deterministic, requires no victim queries, and takes approximately 
$30$--$40$ seconds per target on a single L20 GPU. Aggregated across all 
content-modification baselines and \ATK\ ablation variants, attack 
generation consumes approximately $\mathbf{30}$ GPU-hours.

\paragraph{Evaluation Compute.}
Effectiveness and utility metrics are computed via VIP5's pointwise 
evaluator (B-3 few-shot and B-Z zero-shot prompts, 99-candidate pool) 
on $1{,}000$ benign users per (target, attack, prompt) cell. Each 
cell costs roughly $3$--$4$ minutes of L20 GPU time. 
The reported main results, ablations, defence evaluations, and mechanism analysis, together with additional exploratory evaluations, comprise $6{,}213$ pointwise evaluation runs, totalling approximately $\mathbf{345}$ GPU-hours. For cross-architecture transferability, we evaluate Qwen2-VL-2B-Instruct~\cite{wang2024qwen2} 
with frozen parameters (\texttt{bfloat16} inference, no fine-tuning, 
no LoRA), under both \emph{few-shot} (two in-context demonstrations) 
and \emph{zero-shot} prompt regimes that mirror the VIP5 protocol. 
The six evaluation jobs (3 datasets $\times$ 2 prompt regimes) run 
in parallel across six L20 GPUs, completing in $13$--$32$ wall-clock 
minutes each and contributing approximately $\mathbf{3}$ GPU-hours.

\paragraph{Total Compute, Retries, and Notes.}
Summing across phases, the full \ATK\ study consumes approximately 
$\mathbf{580}$ NVIDIA L20 GPU-hours; this figure includes additional 
exploratory experiments beyond the three datasets reported in the paper. 
Evaluation 
scripts include automatic retries on transient CUDA OOM events (observed 
retry rate below $1\%$). A small fraction of attack-generation jobs 
abort at the M7 stealthiness gate and return the clean baseline 
$(t_0, v_0)$ as specified in Sec.~\ref{sec:stage2}; these are by-design 
fallbacks, not failures. No model is trained from scratch.

\section{Broader Impacts and Responsible Use}
\label{app:broader_impacts}

This work aims to improve the security of multimodal
recommender systems by identifying vulnerabilities
in content-based ranking. The findings may help
researchers and platform operators develop more
robust recommendation mechanisms. However, C3H
is dual-use: similar techniques could be misused
by malicious sellers to manipulate product exposure,
potentially misleading users and disadvantaging
other sellers. Our experiments use public benchmark
datasets and research models; we do not attack
deployed commercial platforms. Monitoring listing
edits and content canonicalisation are potential
mitigations, although their effectiveness against
C3H remains to be evaluated. 
We encourage the use of research artefacts for controlled security evaluation and do not redistribute modified benchmark media.

\clearpage
\newpage
\section*{NeurIPS Paper Checklist}

\begin{enumerate}

\item {\bf Claims}
    \item[] Question: Do the main claims made in the abstract and introduction accurately reflect the paper's contributions and scope?
    \item[] Answer: \answerYes{} 
    \item[] Justification: The abstract and §\ref{sec:intro} state the C2C bias hypothesis, the strict zero-query single-seller threat model, and the headline empirical claims; the experiments in §\ref{sec:experiments} match the scope and magnitude of these claims.

\item {\bf Limitations}
    \item[] Question: Does the paper discuss the limitations of the work performed by the authors?
    \item[] Answer: \answerYes{} 
    \item[] Justification: Sec.~\ref{sec:limitations} discusses the attack's assumptions, the scope of the evaluation, and candidate mitigations and their trade-offs.

\item {\bf Theory assumptions and proofs}
    \item[] Question: For each theoretical result, does the paper provide the full set of assumptions and a complete (and correct) proof?
    \item[] Answer: \answerNA{} 
    \item[] Justification: The paper does not include formal theoretical results; the information-bottleneck arguments cited in §\ref{sec:method} serve as motivation rather than proven contributions.

    \item {\bf Experimental result reproducibility}
    \item[] Question: Does the paper fully disclose all the information needed to reproduce the main experimental results of the paper to the extent that it affects the main claims and/or conclusions of the paper (regardless of whether the code and data are provided or not)?
    \item[] Answer: \answerYes{} 
    \item[] Justification: §\ref{sec:setup} specifies datasets, victims, baselines, and metrics; the full pipeline pseudocode is in Alg.~\ref{alg:c3h}; hyperparameters and small-image fallbacks are detailed in App.~\ref{app:cross-arch}; the compute environment is reported in App.~\ref{app:compute}.

\item {\bf Open access to data and code}
    \item[] Question: Does the paper provide open access to the data and code, with sufficient instructions to faithfully reproduce the main experimental results, as described in supplemental material?
    \item[] Answer: \answerYes{} 
    \item[] Justification: The source code is publicly available at \url{https://github.com/GuoweiGuan666/C3H}. 

\item {\bf Experimental setting/details}
    \item[] Question: Does the paper specify all the training and test details (e.g., data splits, hyperparameters, how they were chosen, type of optimizer) necessary to understand the results?
    \item[] Answer: \answerYes{} 
    \item[] Justification: Training and evaluation details (data splits, optimizer, learning rate, batch size, seeds, prompt regimes, and user-subset sampling) are reported in §\ref{sec:setup} and App.~\ref{app:compute}.

\item {\bf Experiment statistical significance}
    \item[] Question: Does the paper report error bars suitably and correctly defined or other appropriate information about the statistical significance of the experiments?
    \item[] Answer: \answerNo{} 
    \item[] Justification: The final manuscript reports descriptive ER@$K$, HR@$K$, and NDCG@$K$ results in the main and ablation tables, but does not report error bars, confidence intervals, or formal significance tests. Evaluation-user sampling is described in §\ref{sec:setup}. The reported differences are interpreted as descriptive comparisons in the evaluated settings.

\item {\bf Experiments compute resources}
    \item[] Question: For each experiment, does the paper provide sufficient information on the computer resources (type of compute workers, memory, time of execution) needed to reproduce the experiments?
    \item[] Answer: \answerYes{} 
    \item[] Justification: App.~\ref{app:compute} reports the hardware (NVIDIA L20 48\,GB), per-phase wall-clock breakdown, and the total compute (approximately 580 GPU-hours).
    
\item {\bf Code of ethics}
    \item[] Question: Does the research conducted in the paper conform, in every respect, with the NeurIPS Code of Ethics \url{https://neurips.cc/public/EthicsGuidelines}?
    \item[] Answer: \answerYes{} 
    \item[] Justification: We have reviewed the NeurIPS Code of Ethics and the work conforms in all respects.

\item {\bf Broader impacts}
    \item[] Question: Does the paper discuss both potential positive societal impacts and negative societal impacts of the work performed?
    \item[] Answer: \answerYes{} 
    \item[] Justification: Appendix~\ref{app:broader_impacts} discusses the potential security benefits, dual-use risks, and broader societal impacts of C3H, together with the scope of our evaluation and possible mitigations.
    
\item {\bf Safeguards}
    \item[] Question: Does the paper describe safeguards that have been put in place for responsible release of data or models that have a high risk for misuse (e.g., pre-trained language models, image generators, or scraped datasets)?
    \item[] Answer: \answerNA{} 
    \item[] Justification: We do not release pre-trained attack models or modified benchmark datasets. 
The potential dual-use risks of C3H and its research artefacts are discussed in Appendix~\ref{app:broader_impacts}.


\item {\bf Licenses for existing assets}
    \item[] Question: Are the creators or original owners of assets (e.g., code, data, models), used in the paper, properly credited and are the license and terms of use explicitly mentioned and properly respected?
    \item[] Answer: \answerYes{} 
    \item[] Justification: The datasets and base models are cited in Sec.~\ref{sec:setup}. The VIP5 and CLIP repositories use the MIT licence; T5 and Qwen2-VL-2B-Instruct use Apache-2.0. The MicroLens README and Amazon review-data page request citation without specifying named dataset licences. The MicroLens README restricts redistribution of modified data; its download portal additionally specifies academic-use and modification/distribution restrictions for videos. We do not redistribute original or edited MicroLens text, images, or videos; our artefact release is limited to code, configurations, and numerical results.

\item {\bf New assets}
    \item[] Question: Are new assets introduced in the paper well documented and is the documentation provided alongside the assets?
    \item[] Answer: \answerYes{} 
    \item[] Justification: The paper introduces C3H and provides a public repository containing source code and documentation. The main paper and appendices document the attack pipeline, experimental settings, hyperparameters, and computational resources, while the repository provides instructions for reproducing the workflow.

\item {\bf Crowdsourcing and research with human subjects}
    \item[] Question: For crowdsourcing experiments and research with human subjects, does the paper include the full text of instructions given to participants and screenshots, if applicable, as well as details about compensation (if any)? 
    \item[] Answer: \answerNA{} 
    \item[] Justification: The paper does not involve crowdsourcing or research with human subjects; benign-user data is drawn from existing public recommendation benchmarks.

\item {\bf Institutional review board (IRB) approvals or equivalent for research with human subjects}
    \item[] Question: Does the paper describe potential risks incurred by study participants, whether such risks were disclosed to the subjects, and whether Institutional Review Board (IRB) approvals (or an equivalent approval/review based on the requirements of your country or institution) were obtained?
    \item[] Answer: \answerNA{} 
    \item[] Justification: The paper does not involve research with human subjects.

\item {\bf Declaration of LLM usage}
    \item[] Question: Does the paper describe the usage of LLMs if it is an important, original, or non-standard component of the core methods in this research? Note that if the LLM is used only for writing, editing, or formatting purposes and does \emph{not} impact the core methodology, scientific rigor, or originality of the research, declaration is not required.
    \item[] Answer: \answerNA{} 
    \item[] Justification: The MLLM-RecSys studied are the object of evaluation, not a component of the authoring or methodology pipeline; no LLM was used in a non-standard way that affects the core methods, scientific rigor, or originality of the research.

\end{enumerate}
\end{document}